\documentclass[11pt]{article}
\usepackage{amsmath,amssymb,color,graphics,epsfig,cite}

\usepackage{amsfonts}

\newcommand{\hoch}[1]{$\, ^{#1}$}

\newcommand{\be}{\begin{equation}}
\newcommand{\ee}{\end{equation}}
\newcommand{\bea}{\setlength\arraycolsep{2pt} \begin{eqnarray}}
\newcommand{\eea}{\end{eqnarray}}
\newcommand{\nn}{\nonumber}

\def\ft#1#2{{\textstyle{\frac{\scriptstyle #1}{\scriptstyle #2} } }}
\def\fft#1#2{{\frac{#1}{#2}}}

\def\0{{\sst{(0)}}}
\def\1{{\sst{(1)}}}
\def\2{{\sst{(2)}}}
\def\3{{\sst{(3)}}}
\def\4{{\sst{(4)}}}
\def\5{{\sst{(5)}}}
\def\6{{\sst{(6)}}}
\def\7{{\sst{(7)}}}
\def\8{{\sst{(8)}}}
\def\sst#1{{\scriptscriptstyle #1}}

\begin{document}

\begin{center}
{\Large {\bf $D=5$ Rotating Black Holes in Einstein-Gauss-Bonnet Gravity:
Mass and Angular Momentum in Extremality}}

\vspace{20pt}

{\large Liang Ma\hoch{1}, Yue-Zhou Li\hoch{1,2} and H. L\"u\hoch{1}}

\vspace{10pt}

{\it \hoch{1}Center for Joint Quantum Studies and Department of Physics,\\
School of Science, Tianjin University, Tianjin 300350, China \\
{\it \hoch{2} Department of Physics, McGill University, 3600 Rue University, Montr\'{e}al, QC Canada\\
}}

\vspace{40pt}

\underline{ABSTRACT}
\end{center}

We consider perturbative solutions in Einstein gravity with higher-derivative extensions and address some subtle issues of taking extremal limit.   As a concrete new result, we construct the perturbative rotating black hole in five dimensions with equal angular momenta $J$ and general mass $M$ in Einstein-Gauss-Bonnet gravity, up to and including the linear order of the standard Gauss-Bonnet coupling constant $\alpha$.  We obtain the near horizon structure of the near extremal solution, with the blackening factor of the order $\alpha$. In the extremal limit, the mass-angular momentum relation reduces to $M=\frac32 \pi^{\frac13} J^{\frac23} + \pi \alpha$.  The positive sign of the $\alpha$-correction implies that the centrifugal repulsion associated with rotations becomes weaker than the gravitational attraction under the unitary requirement for the Gauss-Bonnet term.

\vfill{\footnotesize  liangma@tju.edu.cn \ \ \ liyuezhou@physics.mcgill.ca \ \ \ mrhonglu@gmail.com}


\thispagestyle{empty}
\pagebreak

\tableofcontents
\addtocontents{toc}{\protect\setcounter{tocdepth}{2}}

\newpage

\section{Introduction}

It is natural to extend Einstein gravity with higher-order curvature invariants.  There are two philosophy regarding these extensions.  One is to treat them as part of a classical theory and consequently the propagators are modified. It was shown that Einstein gravity with the quadratic extension are renormalizable owing to the higher derivative propagator \cite{Stelle:1976gc}.  The price is that the massive spin-2 mode that is responsible to renormalizability is ghostlike.  There exist special combinations, such as the Gauss-Bonnet or the more general Lovelock series \cite{lovelock}, for which the equations of motion remain two derivatives at most on each metric component.  Although they provide no resolution for renormalisability, they could be alternative contenders to Einstein gravity. However, these terms are nontrivial only in higher than four spacetime dimensions. Furthermore, causality analysis from amplitudes put strong constraints on these coupling constants \cite{Camanho:2014apa}.

The alternative view is to treat all the high-derivative terms as part of the low-energy  effective action of a quantum theory of gravity.  In this case, the coupling constants should be sufficiently small and the terms should be treated perturbatively; otherwise, further higher order corrections would have to be all included.  In this approach, the Minkowski spacetime vacuum of Einstein gravity will not be modified by the perturbative correction; furthermore, propagators are unchanged, since all the massive modes are decoupled. Nevertheless, one can learn many important general quantum properties by studying the classical black hole solutions, perturbed by these higher order terms.  It was recently shown \cite{Cheung:2018cwt} that the mass change of the extremal Reissner-Nordstr\"om (RN) black hole under the higher order correction is opposite in sign to the change of entropy, therefore providing a strong evidence of the Weak Gravity Conjecture (WGC) \cite{ArkaniHamed:2006dz}. Many interesting works follow, {\it e.g.,} \cite{Hamada:2018dde,Cheung:2019cwi,Aalsma:2019ryi,Loges:2019jzs,Goon:2019faz,Cremonini:2019wdk,Reall:2019sah}.

In Einstein-Maxwell gravity, the extremal RN black hole symbolizes the balance between gravity and electrostatic repulsion, and appropriate higher order corrections in a UV completed theory are expected to make gravity weaker, under the WGC.  Introducing the Maxwell field makes the higher derivative corrections enormously completed. The independent constraints on causality and unitarity, {\it etc.}, are yet to be comprehensive, although there are already many works in literature \cite{Hamada:2018dde,Chen:2019qvr,Bellazzini:2019xts,Loges:2019jzs}.

In this paper, we would like to focus on the higher-order corrections in pure gravity. The theory is much simpler and may nevertheless be self contained since we expect all the matter fields be consistently truncated. In static configurations, there is no force that can balance gravity and hence no extremal static black hole exists.  However, rotating black holes can be extremal, in which case, the centrifugal repulsion balances the gravitational attraction. It is thus natural to ask whether higher-order corrections make gravity stronger or weaker, comparing to the centrifugal force.  We would also like to know how the entropy of the extremal rotating black hole change under the higher order correction. From the Kaluza-Klein perspective, the distortions of the foliating sphere by rotations give rise to the scalar fields in lower dimensions, and the angular momenta become the electric charges.  For example, electrically-charged static AdS black holes in gauged supergravities can be lifted to become {\it rotating} M-branes or D-branes \cite{Cvetic:1999xp}.

One technique issue immediately arises.  Rotating solutions have much less isometry and the metric depends not only the on radial coordinate, but also the latitude angular coordinates. This makes it difficult to construct these solutions.  In this paper, we consider five dimensional black holes with two equal angular momenta $J$, such that the metric is of cohomogeneity one, depending only on the radial coordinate.  In five dimensions, the Riemann tensor squared is irreducible and thus for our purpose we only need to consider the quadratic curvature extension.  We construct the perturbative rotating black hole for general mass and $J$ and study the effect of the quadratic curvature invariants on the $(M,J)$ relation in the extremal limit.

We are particularly interested in the geometric and thermodynamic properties of the rotating black holes in the near extremal region where the non-extremal factor is of the same order of the higher-order coupling constants.  To understand this subtle limit better, we review in section 2 the RN black hole perturbed by the four-derivative invariants.  These include the Einstein-Maxwell-Gauss-Bonnet (EMGB) gravity that admits exact charged static black holes.  The existence of the exact solution can guide one to take any subtle limit.  In section 3, we present the five-dimensional rotating black holes with equal angular momenta in Einstein-Gauss-Bonnet gravity, up to and including the linear order of the standard Gauss-Bonnet coupling constant.  We study the geometric and thermodynamic properties in the near extremal region.  We conclude the paper in section 4.  In appendix A, we give the general perturbative rotating solution with all six integration constants. In appendix B, we discuss the thermodynamic instability of the RN and Kerr black holes in parallel.

\section{Einstein-Maxwell gravity with four-derivative extensions}

In this section, we digress from our main theme and consider Einstein-Maxwell gravity extended with the general four-derivative invariants built from the Riemann tensor and the Maxwell field strength, for the purpose to understand better the relations of mass, charge and entropy in the extremal limit, in the presence of the higher order corrections.  The most general Lagrangian, up to total derivative terms, is given by
\bea
{\cal L} &=& \sqrt{-g} (R - \ft14 F^2  + \Delta L)\,,\nn\\
\Delta L &=& c_1 R^2 + c_2 R^{\mu\nu} R_{\mu\nu} + c_3 R^{\mu\nu\rho\sigma} R_{\mu\nu\rho\sigma} + c_4 R F^2 + c_5 R^{\mu\nu} F_{\mu\rho} F_\nu{}^\rho\nn\\
&&+ c_6 R^{\mu\nu\rho\sigma} F_{\mu\nu} F_{\rho\sigma} + c_7 (F^2)^2 + c_8 F^{\mu}{}_\nu F^\nu{}_\rho F^\rho{}_\sigma F^{\sigma}{}_\mu + c_9 \nabla_\mu F^{\mu\rho}\nabla^\nu F_{\nu\rho}\,.\label{genlag}
\eea
(We shall not consider the dimension specific topological structures such as $A\wedge F\wedge F$, $A\wedge R\wedge R$, {\it etc}.) In this paper, we consider the case where the coupling constants of the higher-order terms are much smaller than the curvature or field strength invariants of the same dimension so that these term can be treated perturbatively.  This implies that the massive modes are all decoupled and the theory is equivalent under the field redefition\footnote{We are grateful to Jun-Bao Wu for pointing out the $\lambda_5$ redefinition. See also \cite{Liu:2008kt,Cremonini:2009ih,Bueno:2019ltp}.}
\bea
g_{\mu\nu} \rightarrow g_{\mu\nu} + \lambda_1 R_{\mu\nu} + \lambda_2 R g_{\mu\nu} +
\lambda_3 F_{\mu\rho} F_{\nu}{}^\rho + \lambda_4 F^2 g_{\mu\nu}\,,\qquad
A_\mu \rightarrow A_\mu + \lambda_5 \nabla^\nu F_{\nu\mu}\,,\label{redef}
\eea
where $\lambda_i$ are small, of the same order of $c_i$.  Under this redefinition, the coupling constants transform
as
\bea
&&c_{1}  \rightarrow  c_{1}+\ft12\lambda_{1}+\ft12\left(
D-2\right)  \lambda_{2}\,,\qquad
c_{2}  \rightarrow  c_{2}-\lambda_{1}\,,\qquad c_{3}  \rightarrow  c_{3}\,,\nn\\
&&c_{4}\rightarrow  c_{4}-\ft18\lambda_{1}-\ft18\left(
D-4\right)  \lambda_{2}+\ft12\lambda_{3}+\ft12\left(  D-2\right)
\lambda_{4}\,,\nn\\
&&c_{5} \rightarrow c_{5}+\ft12\lambda_{1}-\lambda_{3}\,,\qquad c_{6}  \rightarrow c_{6}\,,\nn\\
&&c_{7}  \rightarrow c_{7}-\ft18\lambda_{3}-\ft18\left(D-4\right) \lambda_{4}\,,\qquad
c_{8}   \rightarrow c_{8}+\ft12\lambda_{3},,\qquad c_9\rightarrow c_9+\lambda_5\,.
\eea
Although there is no invariant structure in the Lagrangian, four special combinations of the coefficients are unmodified by the transformation, and they are \cite{Cheung:2018cwt}
\bea
d_{0}  &=& \ft{1}{2}\left(  D-3\right)  \left(  D-4\right)  ^{2}c_{1}%
+\ft{1}{2}\left(  D-3\right)  \left(  2D^{2}-11D+16\right)  c_{2}\nonumber\\
&&  +\left(  2D^{3}-16D^{2}+45D-44\right)  c_{3}+2\left(  D-2\right)  \left(
D-3\right)  \left(  D-4\right)  c_{4}\nonumber\\
&&  +2\left(  D-2\right)  \left(  D-3\right)  ^{2}\left(  c_{5}+c_{6}\right)
+8\left(  D-2\right)  ^{2}\left(  D-3\right)  \left(  c_{7}+\ft{1}{2}
c_{8}\right),\nn\\
d_1 &=& c_3\,,\qquad d_2 = c_6\,,\qquad d_3 = c_2 + 2 c_5 + 4 c_8\,.
\eea
One can thus use the field redefinition to simplify the four-derivative structures, leaving four inequivalent classes.  We would like to choose these four classes as follows
\be
\Delta L= \alpha L_{\rm GB}+ \beta L_{\rm H}  + \gamma (F^2)^2 + \tilde \gamma L_{\rm qt}\,,\label{DeltaL}
\ee
where $L_{\rm GB}$, $L_{\rm H}$ and $L_{\rm qt}$ are the Gauss-Bonnet, Horndeski \cite{Horndeski:1974wa} and quasi-topological electromagnetism combinations \cite{Liu:2019rib}, given by
\bea
L_{\rm GB} &=& R^2 - 4 R^{\mu\nu} R_{\mu\nu} + R^{\mu\nu\rho\sigma} R_{\mu\nu\rho\sigma}\,,\nn\\
L_{\rm H} &=& R F^2 - 4 R^{\mu\nu} F_{\mu\rho} F_{\nu}{}^\rho + R^{\mu\nu\rho\sigma} F_{\mu\nu} F_{\rho\sigma}\,,\nn\\
L_{\rm qt} &=& (F^2)^2 - 2 F^{\mu}{}_\nu F^\nu{}_\rho F^\rho{}_\sigma F^{\sigma}{}_\mu\,.\label{threestructure}
\eea
(The advantage of these combinations is that ghost excitations can be avoided if one would like to promote these as the starting classical theories, and sometime exact solutions can be constructed, {\it e.g.,} \cite{bdw,cai,Cvetic:2001bk,Feng:2015sbw}.) In terms of these combinations, we have
\bea
&&d_0 = -\ft12 (D-2) \Big((D-4) (3 D-7)\alpha+4(D-3) (2 D-5)\beta-16(D-3) (D-2)\gamma\Big),\nn\\
&&d_1=\alpha\,,\qquad d_2=\beta\,,\qquad d_3=-4 (\alpha +2 \beta +2 \tilde \gamma )\,.\label{d0again}
\eea

\subsection{Einstein-Maxwell-Gauss-Bonnet gravity}
\label{sec:EMGB}

In general, as a low energy effective theory involving the Maxwell field, all terms in (\ref{genlag}), up to the field redefinition (\ref{redef}), should be a priori included, and the coupling constants are constrained by the UV unitarity and analyticity \cite{Adams:2006sv}, {\it etc}. Nevertheless, for simplicity, we consider EMGB gravity as a toy example to gain some insights into the properties of black holes. The advantage of EMGB gravity is that it admits exact solutions of static black holes \cite{bdw,cai,Cvetic:2001bk}.  The Lagrangian is
\be
{\cal L} = \sqrt{-g} \Big(R - \ft14 F^2 + \alpha (R^2 - 4 R^{\mu\nu} R_{\mu\nu} + R^{\mu\nu\rho\sigma} R_{\mu\nu\rho\sigma})\Big).
\ee
For simplicity, we shall consider only  $D=5$ dimensions, where the Gauss-Bonnet term is nontrivial.  The most general ansatz for the spherically-symmetric and static charged black hole is
\be
ds_5^2 = - h(r) dt^2 + \fft{dr^2}{f(r)} + r^2 d\Omega_3^2\,,\qquad A=a(r) dt\,.\label{staticansatz}
\ee
If $\alpha=0$, this leads to the standard RN black hole in five dimensions:
\be
f=1 - \fft{2\mu}{r^2} + \fft{q^2}{r^4}\,,\qquad a=\fft{\sqrt3 q}{r^2}\,.
\ee
The solution is asymptotic to the flat Minkowski spacetime. The mass $M$ and the electric charge $Q$ are defined by
\be
M=\fft{3\pi}{4} \mu\,,\qquad Q=\fft{1}{16\pi} \int {*F}=\fft{\sqrt3 \pi}{4} q\,.\label{masscharge}
\ee
The solution describes a black hole when $\mu\ge q$, with outer and inner horizon radii
\be
r_\pm = \sqrt{\mu \pm \sqrt{\mu ^2-q^2}}\,.
\ee
The first law of black hole thermodynamics $dM=TdS + \Phi dQ$ can be easily established with
\be
S=\ft12 \pi^2 r_+^3\,,\qquad T=\fft{f'(r_+)}{4\pi}\,,\qquad \Phi=\fft{\sqrt3 q}{r_+^2}\,.
\ee
The extremal limit ($\mu=q$) gives the mass-charge relation
\be
M^{\rm ext}=\sqrt3 Q\,.\label{masschargeext}
\ee
Further relevant thermodynamic properties are presented in the appendix \ref{app:stability}. Since $(M,Q)$ are linearly proportional to $(\mu,q)$ respectively, with fixed numerical coefficients, we shall simply use $(\mu,q)$ to denote the mass and charge, unless when we need to obtain the precise first law of black hole thermodynamics in the standard form.

We now consider $\alpha\ne 0$, in which case, $a(r)$ remains unchanged, but the metric functions become \cite{Cvetic:2001bk}
\be
h=f=1 + \frac{r^2}{4 \alpha }\Big(1-\sqrt{1+\frac{16 \alpha  \mu }{r^4}-\frac{8 \alpha q^2}{r^6}}\Big).\nn\\
\ee
Here we chose the branch of the solution that continues to be asymptotically flat. The mass and charge remain the same forms, given by (\ref{masscharge}). The inner and outer horizons are now modified by the Gauss-Bonnet coupling:
\be
r_\pm = \sqrt{(\mu-\alpha) \pm \sqrt{(\mu-\alpha)^2-q^2}}\,.
\ee
The reality condition requires that
\be
\mu\ge q + \alpha\,.
\ee
The temperature, entropy and chemical potential are now given by
\be
T=\fft{r_+^4-q^2}{2\pi r_+^3 (r_+^2 + 4\alpha)}\,,\qquad
S=\ft12 \pi ^2 \left(r_+^3+12 \alpha  r_+ \right)\,,\qquad
\Phi=\fft{\sqrt3 q}{r_+^2}\,.
\ee
In the extremal limit $T^{\rm ext}=0$, we have
\be
\mu^{\rm ext} = q + \alpha\,,\qquad r_+^{\rm ext} = \sqrt{q}\,,\qquad
 \Phi^{\rm ext} =\sqrt3\,,\qquad S^{\rm ext} = \ft12 \pi^2 q^{\fft32} + 6\alpha \pi^2 \sqrt{q}\,.
\ee
One important physical quantity that attracts attention is how the mass-charge relation changes in the extremal limit, which, as we can see above, is particularly simple in this toy model. Note that the Gauss-Bonnet correction has the effect of turning the linear mass-charge relation (\ref{masschargeext}) to subadditivity or superadditivity depending on the sign of $\alpha$:
\bea
&&M(Q_1+Q_2) < M(Q_1) + M(Q_2)\,,\qquad\hbox{for}\qquad \alpha>0\,,\nn\\
&&M(Q_1 + Q_2) = M(Q_1) + M(Q_2)\,,\qquad \hbox{for}\qquad \alpha=0\,,\nn\\
&&M(Q_1+Q_2) > M(Q_1) + M(Q_2)\,,\qquad \hbox{for}\qquad \alpha<0\,.\label{super-sub}
\eea
The subadditivity arising from $\alpha>0$ indicates that charged black holes are energetically favoured for them to join together, a sign of gravity becoming stronger under the Gauss-Bonnet perturbation.  The situation goes opposite when $\alpha<0$. Thus if we treat EGB gravity as some effective field theory, the WGC would necessarily select $\alpha<0$, which is opposite in sign typically discussed in literature.

The above discussion prompts us to study another important physical quantity, the entropy difference due to the higher-order correction. This quantity is more subtle when the charged black hole is close to extremal, but for general mass and charges, we have
\bea
\Delta S &=& S(\mu,q,\alpha) - S(\mu,q,\alpha=0)\label{DSgen}\\
&=& q^{\fft32} \sum_{k=0}^\infty \fft{b_k(\xi)}{\sqrt{\xi}}\Big(\fft{\alpha}{\xi q}\Big)^k,\label{DSgen1}
\eea
where $\xi$ is a dimensionless quantity defined by
\be
\xi = \fft{\mu}{q}-1\,,
\ee
and the coefficients $b_k$ are non-singular function of $\xi$.  We give two leading order coefficients
\bea
b_1 &=& \frac{3 \pi ^2 \sqrt{\xi +\sqrt{\xi } \sqrt{\xi +2}+1} \left(-\xi +7 \sqrt{\xi } \sqrt{\xi +2}-1\right)}{4 \sqrt{\xi +2}}\,,\nn\\
b_2 &=& \frac{3 \pi ^2 \sqrt{\xi +\sqrt{\xi } \sqrt{\xi +2}+1} \left(-15 \xi ^2-30 \xi +(\xi +1) \sqrt{\xi } \sqrt{\xi +2}-2\right)}{16 (\xi +2)^{3/2}}\,.
\eea

For general higher-order corrections, there is no luxury of obtaining the exact solution.  In fact, when higher-order terms are treated as the low energy effective corrections, it is unwarranted to obtain the full solution since the higher-order $\alpha$ corrections are likely modified by even higher-order curvature terms.  At the leading $\alpha$ correction, we have
\bea
\Delta S &=& \frac{3 \pi ^2  \sqrt{\mu +\sqrt{\mu ^2-q^2}} \left(7 \sqrt{\mu ^2-q^2}-\mu \right)}{4 \sqrt{\mu ^2-q^2}}
\alpha + {\cal O}(\alpha^2)\nn\\
&=& \frac{3 \pi ^2 \alpha  \sqrt{\xi +\sqrt{\xi } \sqrt{\xi +2}+1} \left(-\xi +7 \sqrt{\xi } \sqrt{\xi +2}-1\right) \sqrt{q}}{4 \sqrt{\xi } \sqrt{\xi +2}} + {\cal O}\Big(\fft{\alpha^2}{\xi^{3/2}}\Big)\,.\label{alpexp2}
\eea
In fact, at this linear $\alpha$ order, $\Delta S$ can be obtained without solving the equations of motion, but by simply computing the Euclidean action of the higher order curvature terms using the background solution \cite{Reall:2019sah}, based on the quantum static relation (QSR) \cite{Gibbons:1976ue}.\footnote{We would like to cautiously remark that it was shown \cite{Feng:2015oea} that in some black holes of Horndeski gravity, entropy obtained from QSR are not the same as the one obtained from the Wald entropy formula \cite{Wald:1993nt}. The criteria that these two approaches yield the same result remain murky at this stage.}  Since the Euclidean action involves $\beta$, or the inverse of temperature, the statement becomes more subtle and the refinement can be found in \cite{Goon:2019faz}.

In the parameter region
\be
\fft{\alpha}{q} \ll \xi\ll  1\,,\label{xicond1}
\ee
the $\alpha$ expansion remains valid, and furthermore, we can perform small $\xi$ expansion, and this leads to
\be
\Delta S_{\rm CLR} = \fft{3\pi^2\sqrt q}{4}{\alpha}\Big( - \fft{1}{\sqrt{2\xi}} + \fft{13}2 + 3\sqrt{2\xi} + \cdots\Big) + {\cal O}\Big(\fft{\alpha^2}{\xi^{\fft32}}\Big).\label{deltaSliu}
\ee
Thus one sees that the sign of leading $\Delta S_{\rm CLR}$ is opposite to that of $\Delta \mu^{\rm ext}=\alpha$.  In other words, we have $\Delta \mu^{\rm ext} \Delta S_{\rm CLR}<0$.  This turns out to be rather universal for higher-order derivative corrections \cite{Cheung:2018cwt,Goon:2019faz}. Consequently the WGC requirement $\Delta \mu^{\rm ext}<0$ is always followed consistently by $\Delta S_{\rm CLR}>0$.

It is worth noting that the positive condition $\Delta S>0$ established in \cite{Cheung:2018cwt} for thermodynamically stable black hole then implies $\alpha<0$.  However, Gauss-Bonnet coupling $\alpha$ is expected to be positive from some unitary consideration, {\it e.g.,} \cite{Cheung:2016wjt}. This contradiction implies that EMGB on its own is not an appropriate effective field theory, and further corrections in (\ref{genlag}) of the same order should be included. By including all the terms \cite{Cheung:2018cwt}, we have sufficiently enough coupling parameters such that WGC can be restored. As we shall see later, this provides necessarily constraints on a well-defined effective field theory. From an effective field theory analysis, (\ref{xicond1}) is equivalent to \cite{Cheung:2018cwt}
\be
\fft{m_{\rm heav}^3}{\kappa}\ll \xi\ll 1\,,\label{xicond1n}
\ee
where $m_{\rm heav}$ is the mass of heavy-particle that is integrated out in a low energy effective theory, and we restore the Newton constant $\kappa\gg1$ which is defined by
\be
\mathcal{L}=\kappa \sqrt{-g} R+\cdots\,.
\ee

In this paper, we would like to probe the extremality or the near extremal region further, by taking $\xi$ to be in the region of the order $\alpha/q$.  In this case, the naive $\alpha$ expansion (\ref{alpexp2}) is no longer valid.
In fact, we see that $\mu\rightarrow q$ limit is singular in (\ref{alpexp2}). This singularity is clearly an artifact of the $\alpha$ expansion while holding $\mu\gg q$ fixed, since $\Delta S$ defined in (\ref{DSgen}) is clearly regular in the $\mu\rightarrow q$ limit. Furthermore, when $\mu-q\sim \alpha$, every term in (\ref{DSgen1}) becomes of the same order, and the expansion scheme becomes invalid and a new perturbative parameter is called for.

As we saw earlier, in the extremal limit, $\mu^{\rm ext}=q + \alpha$.  This implies that for given mass and charge $(\mu,q)$, the RN and the $\alpha$-corrected solutions cannot be simultaneously extremal. We thus need to define $\Delta S$ properly in the region of extremality.  In one definition, we may simply hold the charge fixed, and obtain the extremal solutions for both $\alpha=0$ and $\alpha\ne0$ and then compare the difference, namely
\be
\Delta_0 S = S^{\rm ext}(q,\alpha) - S^{\rm ext}(q,\alpha=0)\,.
\ee
In this definition, the mass of the two extremal solutions are in general not the same; exceptions exist in supersymmetric theories \cite{Cano:2019ycn}. If we insist on the entropy difference for the same mass and charge, two situations can arise. For $\alpha>0$, we have $\mu^{\rm ext}=q + \alpha>q$, it follows that when we can take extremal limit for the $\alpha$-corrected solution, the original RN black hole of the same mass and charge becomes near extremal, whose blackening factor is of the order $\alpha$.  We can thus define
\be
\Delta_+ S = S^{\rm ext}(q, \alpha) - S(\mu=q+\alpha, q, \alpha=0)\,,\qquad \hbox{for}\qquad \alpha>0\,.
\ee
When $\alpha<0$, the RN black hole of mass and charge $(\mu^{\rm ext}=q + \alpha, q)$ ceases to be a black hole.  It is thus more natural to define the entropy difference by
\be
\Delta_- S = S(\mu=q,q,\alpha) - S^{\rm ext}(q,\alpha=0)\,,\qquad \hbox{for}\qquad\alpha<0\,.
\ee
Using the thermodynamic quantities of general $\alpha$ obtained earlier, we find
\bea
\Delta_0 S  &=& 6 \pi ^2 \alpha  \sqrt{q}\,, \nn\\
\Delta_\pm S &=&\mp\frac{3 \pi ^2 \sqrt{\pm\alpha } q}{2 \sqrt{2}}+ \frac{39}{8} \pi ^2 \alpha  \sqrt{q} + {\cal O}((\pm\alpha)^{\fft32})\,.\label{deltaSs}
\eea
This is the furthest way that we can push towards extremality, and we shall discuss the near extremal region of the $\alpha$'th order presently. Incidentally, the sign of the leading order of $\Delta_\pm S$ coincides with the one of $\Delta S_{\rm CLR}$ given in (\ref{deltaSliu}).  However, now we have
\be
\Delta M = - \sqrt{\pm\alpha}\, \fft{\sqrt3}{8Q}\Delta_\pm S + {\cal O}(\alpha^\fft32)\,.
\ee
We see that the condition $\Delta M \Delta_\pm S<0$ continues to hold in the furthest limit to the extremality.
It should be emphasized however while the positive condition $\Delta S>0$ was shown successfully in the region (\ref{xicond1n}) by the effective field theory arguments, there is no analogous demonstration that this must be true when probing further into the extremality, since as $\xi$ goes beyond the regime (\ref{xicond1n}), the effective field theory power-counted by $\alpha$ is no longer valid, and the assumptions that the positive condition was built upon in \cite{Cheung:2018cwt} breaks down.  Nevertheless, the fact $\Delta M \Delta_\pm S<0$, together with the WGC, provides the circumstance evidence that $\Delta S>0$ continues to be true and $\Delta_-S$ should thus be selected. Indeed this branch was chosen in \cite{Hamada:2018dde} where the more general four-derivative theory was studied.

We now show that these results can be obtained from the perturbative solution
\be
h=f=1 - \fft{2\mu}{r^2} + \fft{q^2}{r^4} + \alpha \frac{2\left(q^2-2 \mu  r^2\right)^2}{r^{10}} + {\cal O}(\alpha^2)\,.
\label{emgblinearsol}
\ee
For mass and charge satisfying $\mu -q \gg \alpha$, we have
\be
r_+ = \sqrt{\mu + \sqrt{\mu^2-q^2}} \Big(1 - \fft{\alpha}{2\sqrt{\mu^2-q^2}}\Big) + {\cal O}(\alpha^2)\,.
\ee
This $\alpha$ expansion however is no longer valid as $\mu-q\sim \alpha$ and there is a one-parameter family of $\alpha$-corrected extremal $(\mu=q)$ RN black hole at the linear $\alpha$ order, with the mass and charge relation
\be
\mu = q + (\eta + 1) \alpha\,,\label{nearext0}
\ee
where $\eta$ is a dimensionless order one parameter, satisfying $\eta \alpha \ge 0$, so that we have $\mu \ge q + \alpha$, a condition for the solution to be a black hole after the $\alpha$ correction. Thus $\eta$ can be viewed as a blackening parameter of the near-extremal solutions whose deviation from extremality is of the order $\alpha$, as can be see in (\ref{nearext0}).  The horizon radius expansion now involves in general not only $\alpha$, but $\sqrt{\alpha}$ as the leading order,  given by
\be
r_+=\sqrt{q} + \sqrt{\ft12\alpha \eta} + \fft{\alpha \eta}{4\sqrt q} + {\cal O}((\eta\alpha)^{2})\,.
\ee
Note that we cannot obtain higher order corrections, using the linear perturbative solution (\ref{emgblinearsol}).
The extremal limit is uniquely determined by setting $\eta=0$, and the horizon radius remains $r_+=\sqrt{q}$, unmodified by the higher derivative terms at the linear order.  On the other hand, the extremal $\mu=q$ RN black hole, corresponding to $\eta=-1$, remains a black hole only when $\alpha<0$ and the leading shift of the horizon radius is of order $\sqrt{\alpha}$. With these, it is then straightforward to obtain all the $\Delta S$ given in (\ref{deltaSs}).
We find that the relation
\be
\Delta M \propto -\sqrt{|\alpha|} \Delta S\label{dmds}
\ee
holds for the general near extremal solutions, where $\Delta S$ is understood to be analogous to $\Delta_\pm S$, depending on whether $\alpha$ is positive or negative.  It should be pointed out that the result (\ref{dmds}) is consistent with the statement $\Delta M \propto -T \Delta S$ of \cite{Goon:2019faz} for the near extremal solutions; our result takes the near extremal region further with $T\sim \sqrt{|\alpha|}$.

\subsection{General four-derivative corrections}

In this subsection, we present the full perturbative solution in general $D$ dimensions for the Lagrangian  (\ref{genlag}). We shall concentrate only on the purely electric solution. It follows that the quasi-topological electromagnetism structure $L_{\rm qt}$ in (\ref{threestructure}) does not contribute to the equations of motion \cite{Liu:2019rib}. Since $\nabla_\mu F^{\mu\nu}$ vanishes at the leading order, $\nabla_\mu F^{\mu\rho}\nabla^\nu F_{\nu\rho}$ does not contribute either \cite{Cheung:2018cwt}.  The general ansatz takes the same form as (\ref{staticansatz}), but with $d\Omega_3^2$ replaced by $d\Omega_{D-2}^2$, the metric for the unit round $S^{D-2}$.  The solution is given by
\be
h= f_0 + \Delta h\,,\qquad f= f_0 + \Delta f\,,\qquad a= a_0 + \Delta a\,,
\ee
where
\be
f_0 = 1 - \fft{2\mu}{r^{D-3}} + \fft{q^2}{r^{2(D-3)}}\,,\qquad a_0=\sqrt{\ft{2(D-2)}{D-3}} \fft{q}{r^{D-3}}\,,
\ee
and
\bea
\Delta h &=& \fft{2(D-3)}{(D-2) r^{2(D-2)}}\Bigg(2(D-4)(D-2) \mu^2 c_3 + q^2
\Big(2 (D-4)c_1-\left(D^2-6 D+10\right)c_2\nn\\
&&-2 \left(2 D^2-11 D+16\right)c_3+4 (D-2)c_4-(D-4) (D-2)c_5\cr
&&-2(D-3) (D-2)c_6\Big)\Bigg)-\frac{4 (D-3) \mu  q^2}{(D-2)r^{3D-7}}\Big(
(D-4) (D-1)c_1-c_2- D c_3\nn\\
&&+2(D-2) (D-1)c_4+(D-2)c_5- (D-3) (D-2)c_6\Big)\nn\\
&&+\frac{(D-3) q^4}{(D-2) (3 D-7)r^{2(2D-5)}}\Big((D-4) \left(11 D^2-45 D+44\right)c_1\nn\\
&&+ \left(4 D^3-33 D^2+83 D-64\right)c_2 +2\left(4 D^3-34 D^2+87 D-68\right)c_3\nn\\
&&+4 (D-2) \left(5 D^2-19 D+16\right)c_4 +2 (D-2) \left(D^2-D-4\right)c_5 \nn\\
&&-4  (D-2) (D-3)^2 c_6-16  (D-2)^2 (D-3) c_7-8  (D-2)^2 (D-3)c_8\Big),\cr
\Delta f &=& \Delta h - \frac{4 (D-3) q^2 f_0}{(D-2) r^{2(D-2)}}\Big((D-4) (2 D-3)c_1 +\left(D^2-5 D+5\right)
c_2\nn\\
&&+\left(2 D^2-9 D+8\right)c_3+2 (D-2) (2 D-3)c_4
+(D-2) (D-1)(c_5+c_6)\Big),\nn\\
\Delta a&=& \fft{4 \sqrt{2} c_6 (D-3)^{3/2} \sqrt{D-2} \mu q}{r^{2(D-2)}} +
\frac{2 \sqrt{2} (D-3)^{3/2} q^3}{\sqrt{D-2} (3 D-7) r^{3D-7}}\Big((D-4) (2 D-3)c_1\nn\\
&&+\left(D^2-5 D+5\right)c_2+ \left(2 D^2-9 D+8\right)c_3+2(D+1) (D-2)c_4\nn\\
&&-(D-5) (D-2)c_5- (7 D-19) (D-2)c_6 -8 (D-2)^2(2c_7-c_8)\Big).
\eea
In this perturbation, the mass and charge remain unchanged by the higher-order corrections, namely
\be
M=\frac{\left(  D-2\right)  \Omega_{D-2}}{8\pi}\,\mu\,,\qquad Q=\frac
{\sqrt{2\left(  D-2\right)  \left(  D-3\right)  }\Omega_{D-2}}{16\pi}\,q\,,
\ee
where $\Omega_{D-2}$ is the volume of the round unit $S^{D-2}$. The outer horizon radius of the original unperturbed RN black hole, which we denote as $r_0$, is given by
\be
r_0^{D-3} = \sqrt{\mu + \sqrt{\mu^2-q^2}}\,.
\ee
For fixed $(\mu,q)$ with $\mu-q\gg c_i q^{(D-5)/(D-3)}$, it is straightforward to derive the shifted radius of the outer horizon,
\be
r_+=(\mu + \sqrt{\mu^2-q^2})^{\fft{1}{D-3}} + \Delta r\,.
\ee
One can then obtain the higher-order corrected thermodynamic variables, including the entropy
\bea
S &=& \ft14 \Omega_{D-2} r_+^{D-2}\Bigg(1 -\frac{(2 c_6+c_5+2c_4) f \psi '^2}{h}+\frac{c_3 \left(f \left(h'^2-2 h h''\right)-h f' h'\right)}{h^2}\nn\\
&&-\frac{c_2}{2 h^2 r} \Big(f \left((D-2) h h'+2 r h h''-r h'^2\right)+h f' \left((D-2) h+r h'\right)\Big)\nn\\
&&-\frac{c_1}{h^2 r^2}\Big(rh f' \left(2 (D-2) h+r h'\right)+2 h \left((D-3) (D-2) (f-1) h+r^2f h''\right)\cr
&&+2 (D-2) r f h  h'-r^2f h'^2\Big)\Bigg)\Bigg|_{r=r_+}\nn\\
&=& \ft14 \Omega_{D-2} \Bigg(r_+^{D-2} +\frac{4 c_3 (D-3) (D-2) \mu }{r_+} -
\fft{2(D-3)q^2}{r_+^{D-2}} \Big((D-4)c_1\nn\\
&&+(D-3)c_2+2 (2 D-5)c_3+2  (D-2)c_4+ (D-2)c_5+2 (D-2)c_6\Big)\Bigg) + {\cal O}(c_i^2).
\eea
The $\Delta S_{\rm CLR}$ was obtained in \cite{Cheung:2018cwt} and we shall not repeat here.

Focusing on the perturbation around the extremal $\mu=q$ solution, we find there is one-parameter family of near extremal solutions of order $c_i$,
\bea
\mu &=& q - \fft{(D-3)r_0^{D-5}}{(D-2)(3D-7)} d_0 (1 + \eta)\,,\nn\\
r_+ &=& r_0 + \sqrt{\fft{2\eta d_0}{(D-3)(D-2)(3D-7)}} -
\fft{\eta d_0}{(D-3)(D-2)(3D-7)r_0}\nn\\
&&+ \frac{D-4}{2 (D-3) (D-2) r_0} \Bigg((D-3) (3 D-8)c_1+ (D-3) (2 D-5)c_2\nn\\
&&+2\left(2 D^2-10 D+13\right)c_3\Bigg)+\fft{2(D-2)}{r_0} \left(c_4-4 c_7-2 c_8\right).
\eea
Here $\eta$ is a dimensionless non-extremal parameter of order one, satisfying $\eta d_0\le 0$ for the perturbative solution to be a black hole. The extremality saturates the inequality.  The entropy of the near extremal black holes is
\bea
S&=&\ft14 \Omega_{D-2} r_0^{D-2}\Big(1 + \sqrt{-\frac{2(D-2) \eta d_0}{(D-3) (3 D-7)r_0^2}}
-\fft{(D-2)\eta d_0}{(D-3)(3D-7)r_0^2}\nn\\
&&+\frac{2 (D-3) (D-2) \left(d_1-d_2 (D-3)\right)-d_0}{(D-3) r_0^2}\Big).
\eea
We therefore obtain
\bea
\Delta_0 S &=& \frac{\Omega _{D-2}}{4 (D-3)} r_0^{D-4}\Big(2 (D-3) (D-2) \left(d_1-d_2 (D-3)\right)-d_0\Big),\nn\\
\Delta_\pm S &=& \frac{\Omega _{D-2} r_0^{D-4}}{4 (D-3) (3 D-7)}\Big(\mp
\sqrt{\mp 2 (D-3) (D-2) (3 D-7)d_0}\, r_0 -  (2 D-5)d_0\nn\\
&&+2 (D-3) (D-2) (3 D-7) \left(d_1- (D-3)d_2\right)\Big).
\eea
The result up to the $\sqrt{|d_0|}$ order was obtained in \cite{Hamada:2018dde}. Note that when specialising the parameters to be those of EMGB gravity, we recover (\ref{deltaSs}).  The relation (\ref{dmds}) holds, but with $|\alpha|$ replaced by $|d_0|$.  It should be understood that these quantities are correct up to and including the linear order of the coupling constants and we shall not always be pedantic on writing out all the ${\cal O}(\alpha^2)$ or ${\cal O}(c_i^2)$ expressions.

As discussed earlier, the field redefinition (\ref{redef}) allows us to consider the simpler theory (\ref{DeltaL}) with
$d_0$ given by (\ref{d0again}).  The requirement that $\Delta S_{\rm CLR}$ must be nonnegative implies that $d_0\ge 0$ \cite{Cheung:2018cwt}.  As we have seen earlier, the unitarity requirement for the Gauss-Bonnet coupling constant $\alpha$ gives negative contribution to $d_0\ge 0$.  Analogously the unitarity requires that the Horndeski coefficient $\beta$ must be nonnegative \cite{Hamada:2018dde}, also giving negative contribution to $d_0\ge 0$.  This makes the $\gamma$ term indispensable. The $\gamma$ term involves purely matter and we can impose the standard energy condition.  Fortunately we find that both the strong or dominant energy conditions require that $\gamma\ge 0$, which is also consistent with the unitarity \cite{Hamada:2018dde}. Thus the matter contribution to the higher-order correction is crucial and it should be sufficiently large in order to make it possible to satisfy the constraint $d_0\ge 0$.

Our analysis suggests that the unitary higher-derivative perturbations involving curvature tensor tend to make gravity stronger whilst the pure matter sector tends to make gravity weaker.  We would like to keep this observation in mind in discussing rotating black holes in pure gravity in the next section.

\section{Rotating black holes in pure quadratically extended gravity}
\label{sec:rot}

As we have seen in the previous section, even though Einstein-Maxwell gravity is a fairly simple theory, its next order correction is already quite complicated involving four different inequivalent combinations of 9 invariant structures.
One may naively drop out some higher derivative terms involving the Maxwell field, for example, to keep only the Gauss-Bonnet term as we discussed in the previous section; however, such simplification is illegal in the context of effective field theory and may lead to contradictions with unitarity. Nevertheless, in this section, we would like to consider the simpler theory, by decoupling all the matter fields.  The theory is Einstein gravity extended by the Riemann tensor quadratics
\be
{\cal L} = \sqrt{-g} \Big(R + \alpha R^{abcd} R_{abcd} + \beta R^{ab}R_{ab} + \gamma R^2\Big).\label{RiemannE}
\ee
In this approach, we made an assumption that matter fields can be consistently truncated. Such an assumption is a necessary starting point before we have a compete theory of quantum gravity; in fact, in Einstein-Maxwell gravity discussed in the previous section, we also assumed for example that the scalar fields are all decoupled. As a pure gravity theory with all matter fields decoupled, (\ref{RiemannE}) enumerates all invariant structures with fourth derivatives and is thus legal as an effective field theory. The issue is then whether this theory is relevant for the discussion on the WGC. There is no static extremal solution since there is no repulsion in the static spacetime. However, rotating black holes can be extremal, where the centrifugal force, analogous to the electrostatic repulsion, balances the gravitational attraction.  In fact from the Kaluza-Klein perspective, the angular momentum can be viewed as electric charges in the lower dimensions.  For example, rotating M-branes or D3-branes are directly related  to the $R$-charged black holes in gauged supergravities \cite{Cvetic:1999xp}.  The Kaluza-Klein reduction of the Gauss-Bonnet term keeping both the scalar and $U(1)$ vector field was given in the appendix of \cite{Liu:2012ed}.

In Einstein gravity, the rotating black hole is described by the Kerr metric, constructed in 1963 \cite{Kerr:1963ud}.  The cosmological constant was later introduced \cite{Carter:1968ks}. Motivated by the string development in 80s, the asymptotically flat Kerr black hole was later generalized to all higher dimensions \cite{Myers:1986un}.  After the AdS/CFT correspondence was proposed, Kerr-AdS black hole in five dimensions was quickly constructed \cite{Hawking:1998kw}, and it was later generalized to all higher dimensions \cite{Gibbons:2004js,Gibbons:2004uw}, followed by the general higher-dimensional Kerr-AdS-NUT Plebanski metrics \cite{Chen:2006xh}.  Higher-dimensional Diemnieski-Plebanski metrics have so far been elusive except in $D=5$ and it is asymptotic flat \cite{Lu:2008js}, which interestingly contains the famous black ring solution \cite{Emparan:2001wn}.

Typically rotating black holes are of higher-cohomogeneity, depending on not only the radial coordinate, but also the latitude angle coordinates.  This makes it difficult to construct these metrics, even in Einstein gravity, let alone higher order gravities.  Many constructions in literature use the numerical approach or take the slow rotation approximation, {\it e.g.,} \cite{Brihaye:2008kh,Brihaye:2010wx,Kleihaus:2011tg,Yue:2011zza, Okounkova:2019zep,%
Cano:2019ozf,Adair:2020vso,Konoplya:2020fbx}, which is of no use for our purpose of comparing the centrifugal and gravitational forces in the extremal limit.

\subsection{$D=5$ rotating black hole with equal angular momenta}

In this section, we consider $D=5$, where the quadratic extensions are nontrivial for the Ricci-flat backgrounds.
Furthermore, in five dimensions, there are two perpendicular rotations and the metric reduces to cohomogeneity one when the two rotations are equal. The foliating $S^3$ becomes squashed, preserving $U(1)\ltimes SU(2)$ isometry.  We follow \cite{Feng:2016dbw} and write the most general metric ansatz, involving four functions
\be
ds_5^2 = \fft{dr^2}{f(r)} - \fft{h(r)}{W(r)} dt^2 + \ft14 r^2 W(r) (d\psi + \cos\theta d\phi + \omega(r) dt)^2 +
\ft14 r^2 (d\theta^2 + \sin^2\theta d\phi^2)\,.
\ee
The general five dimensional Kerr metric \cite{Myers:1986un} reduces to the one with equal angular momenta, with
\be
h=f=f_0\equiv 1 - \fft{2\mu}{r^2} + \fft{\nu^2}{r^4}\,,\qquad W=W_0\equiv 1 + \fft{\nu^2}{r^4}\,,\qquad
\omega=\omega_0\equiv \fft{2\sqrt{2\mu}\,\nu}{r^4 W_0}\,.
\ee
The parameter $(\mu,\nu)$ parameterized the mass and angular momentum
\be
M=\ft34\pi \mu\,,\qquad J=\ft14 \pi \sqrt{2\mu}\, \nu\,.
\ee
The reason to cast the Kerr metric in this form in \cite{Feng:2016dbw} is that when the mass is negative ($\mu<0$ and $\nu^2<0$), the metric describes a smooth time machine with the velocity of light surface located at $W_0=0$.  In this paper, we focus on positive mass with $(\mu,\nu)$ both positive.  The solution describes a rotating black hole provided that $\mu\ge \nu$, with the inner and outer horizons
\be
r_\pm^0 = \sqrt{\mu \pm \tilde \nu}\,,\qquad \tilde \nu=\sqrt{\mu^2 - \nu^2}\,.\label{bghorizons}
\ee
The temperature, entropy and angular velocity associated with the outer horizon are
\be
T=\fft{\tilde \nu(\mu - \tilde \nu)}{\pi\sqrt{2\mu} \nu^2}\,,\qquad
S=\pi^2 \sqrt{\ft12\mu} (\mu + \tilde \nu)\,,\qquad
\Omega=\fft{\sqrt2(\mu - \tilde \nu)}{\sqrt{\mu}\nu}
\ee
It is easy to verify that the first law $dM = TdS + \Omega dJ$ is satisfied.  The solution becomes extremal with $T=0$ when $\mu=\nu$, or $\tilde \nu=0$, implying
\be
M^{\rm ext}=\ft{3}{2} \pi^{\fft13} J^{\fft23}\,,\qquad S^{\rm ext} = 2\pi J\,.\label{kerrmassext}
\ee
It is of interest to note that the entropy at the extremality is linearly proportional to the angular momentum. The mass-charge relation however satisfies subadditivity
\be
M(J_1+J_2)\le M(J_1) + M(J_2)\,.\label{subadd}
\ee
By contrast, the mass-charge relation for the extremal RN black is linear and the entropy is a superadditive function of the charge. This implies that the centrifugal repulsion associated with the spin of the black holes is weaker than gravity and two extremal rotating black holes are energetically favoured to join together. We would like to exam next how the quantum effects would modify this. In appendix \ref{app:stability}, we discuss the thermodynamic instability of the rotating black hole, in parallel to the RN black hole.

\subsection{Higher-order corrections}

We are now in the position to consider higher-order corrections to the solution. The non-perturbative solution was constructed numerically in \cite{Brihaye:2008kh}. For our purpose, we would like to construct perturbative analytic solutions. Since the leading Kerr metric is Ricci-flat, the $R^2$ and $R^{ab}R_{ab}$ terms will not contribute to the equations of motion at the linear order,
thus the perturbed solution is also for the Einstein-Gauss-Bonnet gravity with the Gauss-Bonnet coupling $\alpha$.
We therefore consider
\be
h=f_0 + \alpha \Delta h\,,\qquad f=f_0 + \alpha \Delta f\,,\qquad W=W_0 + \alpha \Delta W\,,\qquad
\omega=\omega_0 + \alpha \Delta \omega\,.\label{rotpert}
\ee
We find that the linear perturbative equations can be fully solved for general $(\mu,\nu)$, involving six integration constants. We present the full solutions in appendix \ref{app:sol}.  Choosing the parameters appropriately, we find
\bea
\Delta h &=& \frac{64 \mu  \left(2 \mu ^2+\nu ^2\right) \left(\mu -\tilde{\nu }\right)}{3 \nu ^4 r^2}
-\frac{32 \mu  \left(8 \mu ^2-6 \mu  \tilde{\nu }+\nu ^2\right)}{3 \nu ^2 r^4}+\frac{8 \left(9 \mu ^2+4 \nu ^2\right)}{3 r^6}+\frac{32 \mu  \nu ^2}{3 r^8}\nn\\
&&-\frac{16 \nu ^4}{3 r^{10}} -\fft{64\mu}{3\nu^4 r^4}
\Big(2\mu r^2\left(\mu ^2+2 \nu ^2\right) -\left(2 \mu ^2+\nu ^2\right) \left(\nu ^2+r^4\right)\Big)\log \left(1-\frac{\mu-\tilde \nu }{r^2}\right),\nn\\
\Delta f &=& \frac{64 \mu ^2 \left(\mu -\tilde{\nu }\right)}{\nu ^4}-\frac{32 \mu ^2 \left(2 \mu ^2-2 \mu  \tilde{\nu }+\nu ^2\right)}{\nu ^4 r^2}+\frac{64 \mu ^3}{3 \nu ^2 r^4}+\frac{40 \mu ^2}{3 r^6}+\frac{64 \mu  \nu ^2}{3 r^8}-\frac{16 \nu ^4}{r^{10}}\nn\\
&&+\Big(\frac{64 \mu ^2 r^2}{\nu ^4}-\frac{128 \mu ^3}{\nu ^4}+\frac{64 \mu ^2 \left(2 \mu ^2+\nu ^2\right)}{3 \nu ^4 r^2}\Big)\log \left(1-\frac{\mu-\tilde \nu }{r^2}\right),\nn\\
\Delta W &=& -\frac{32 \mu ^2 \left(\mu -\tilde{\nu }\right)}{\nu ^4}+\frac{16 \mu ^2}{\nu ^2 r^2}
+\frac{4 \left(16 \mu ^4+15 \mu ^2 \nu ^2-6 \nu ^4-8 \mu  \tilde{\nu } \left(2 \mu ^2+\nu ^2\right)\right)}{3 \mu  \nu ^2 r^4}\nn\\
&&-\frac{16 \left(\mu ^2+3 \nu ^2\right)}{3 r^6}-\frac{16 \mu  \nu ^2}{3 r^8} +\frac{16 \nu ^4}{3 r^{10}}\nn\\
&&-32\mu\Big(\frac{\mu r^2}{\nu ^4}-\frac{\mu ^2}{\nu ^4}-\frac{\mu}{\nu ^2 r^2}+\frac{\left(\mu ^2+2 \nu ^2\right)}{3 \nu ^2 r^4}\Big)\log \left(1-\frac{\mu-\tilde \nu }{r^2}\right),\nn\\
\Delta\omega &=& \frac{16 \sqrt{2} \sqrt{\mu } \left(-6 \mu  r^6 \left(r^2-\mu \right) \left(\mu -\tilde{\nu }\right)+2 \nu ^6-2 \mu  \nu ^4 r^2+\mu  \nu ^2 r^4 \left(3 r^2-2 \mu \right)\right)}{3 \nu ^3 r^6 \left(\nu ^2+r^4\right)}\nn\\
&& -\frac{32 \sqrt{2} \mu ^{3/2} \left(2 \mu ^2+\nu ^2+3 r^4-6 \mu  r^2\right)}{3 \nu ^3 \left(\nu ^2+r^4\right)}
\log \left(1-\frac{\mu-\tilde \nu }{r^2}\right).
\eea
Assuming that there exists a horizon $r_+$, the Wald entropy formula \cite{Wald:1993nt} gives
\be
S=\frac{1}{2} \pi ^2 r_+ \sqrt{\nu ^2+r_+^4}+\frac{\pi ^2 \alpha  \left(r_+^8 \Delta W(r_+) +48 \mu  r_+^4 -16 \mu  \nu ^2\right)}{4 r_+^3 \sqrt{\nu ^2+r_+^4}}\,.
\ee

We now consider the perturbation that the mass and angular momentum are fixed, in which case, the horizon is shifted by
\bea
r_+ &=& \sqrt{\mu + \tilde \nu} + \fft{\alpha}{3\mu\nu^2 \tilde \nu (\mu + \tilde \nu)^{\fft72}}\Big(\mu( 32\mu ^6-72 \mu ^4 \nu ^2+51 \mu ^2 \nu ^4-14 \nu ^6)\nn\\
&&\qquad\qquad+\tilde{\nu } (32 \mu ^6 -56 \mu ^4 \nu ^2+27 \mu ^2 \nu ^4-6 \nu ^6)\Big)\nn\\
&&+ \fft{16\alpha\mu \tilde \nu^2}{3\nu^4 (\mu + \tilde \nu)^{\fft72}}\Big(8 \mu ^4-8 \mu ^2 \nu ^2+\nu ^4+4 \mu  \tilde\nu \left(2 \mu^2-\nu^2\right)\Big)\log \left(\frac{2 \tilde{\nu }}{\tilde{\nu }+\mu }\right).
\eea
Note that we only present the outer horizon.  The situation for the inner horizon is more complicated, since the spacetime singularity is not only located at $r=0$, but also at $r=r_0^-$, the inner horizon of the unperturbed solution. This makes the solution very different from the previous example. (The solution whose singularity is located at $r=r_0^+$ is presented in appendix \ref{app:sol}.) The sign choice of this divergent term is crucial for the singularity structure.  Furthermore, for $r_-$ to be the inner horizon, the functions $f(r_-)$ and $h(r_-)$ need to vanish simultaneously. This is difficult to satisfy once we made it so for the outer horizon. For $\alpha<0$, both $h$ and $f$ diverges negatively at the original inner horizon $r_0^-$ and the roots of $h$ and $f$ are all shifted to be bigger than the unperturbed inner and outer horizons. The divergence at $r_0^-$ splits the smaller roots of $h$ and $f$ that are just outside $r_0^-$, but they become closer and closer towards extremality. When $\alpha>0$, both $h$ and $f$ diverges positively and therefore the smaller roots of $h$ or $f$ do not exist at all if $\tilde \nu \gg \alpha$.  As $\tilde \nu$ approaches $\alpha$, two roots develops just outside $r_0^-$, it follows that $h$ and $f$ each can have a total of three real roots and black holes with two horizons can exist at the linear $\alpha$ order. We shall give presently the near-horizon geometry of the near-extremal solution.

The mass, entropy and angular velocity for the general non-extremal solutions far away from extremality become
\bea
T&=&T_0\Big(1 + \frac{2 \alpha  \left(\tilde{\nu } \left(14 \mu ^5-25 \mu ^3 \nu ^2+14 \mu  \nu ^4\right)-14 \mu ^6+32 \mu ^4 \nu ^2-27 \mu ^2 \nu ^4+6 \nu ^6\right)}{3 \mu  \nu ^4 \tilde{\nu }^2}\Big),\nn\\
\Omega &=& \Omega_0 \Big(1+\frac{2 \alpha  \left(\tilde{\nu } \left(8 \mu ^5-19 \mu ^3 \nu ^2+14 \mu  \nu ^4\right)-8 \mu ^6+23 \mu ^4 \nu ^2-21 \mu ^2 \nu ^4+6 \nu ^6\right)}{3 \mu  \nu ^4 \tilde{\nu }^2}
\Big),\nn\\
S &=& S_0 + \Delta S\,,
\eea
where
\be
\Delta S=\frac{\pi ^2 \alpha \sqrt{\mu } \left(9 \left(\tilde{\nu }+\mu \right)^4-14 \nu ^2 \left(\tilde{\nu }+\mu \right)^2+\nu ^4\right)}{2 \sqrt{2} \tilde{\nu } \left(\tilde{\nu }+\mu \right)^3}\,.
\ee
Thus it is clear that in the limit where $|\alpha| \ll \tilde \nu\ll \mu$, the leading order of $\Delta S$ is
$-\sqrt2 \pi^2 \alpha/\tilde \nu$ which is positive only when $\alpha$ is negative, which violates unitarity \cite{Cheung:2016wjt}. We shall comment on this at the end of this section.

We are interested in taking extremal or near extremal limits of the order $\alpha$.  When $\alpha=0$, the extremal solution occurs at $\mu=\nu=r_0^2$, where $r_0$ is the double horizon.  For $\alpha\ne 0$, we would like to construct near extremal solutions with the angular momentum $J$ held fixed.  This implies the following corrections to $\mu$ and $\nu$ parameters
\be
\mu=r_0^2 + \ft43 (1+ \eta) \alpha\,,\qquad \nu = r_0^2 - \ft23(1 + \eta)\alpha\,,
\ee
such that
\be
M=\ft{3}4\pi \Big(r_0^2 + \ft{4}{3} \alpha  (\eta +1)\Big),\qquad
J=\frac{\pi  r_0^3}{2 \sqrt{2}}\,.
\ee
In other words, here $r_0$ is simply a parameter for the angular momentum $J$ and the near extremal factor is of the order $\alpha$, namely $\delta M\sim \eta \alpha$. As remarked earlier, the solution sufficiently far away from the extremality do not have two horizons. However, when $\delta M\sim \alpha$, the two horizon solution exists, and
the inner and outer horizons of the near-extremal solution are
\bea
r_\pm= r_0\pm \sqrt{\alpha  \eta }+\frac{\alpha  (16+\eta)}{6 r_0}\,.
\eea
The near horizon geometry of the near-extremal black hole is given by
\bea
h&=&\left(-\frac{12 r}{r_0^3} +\frac{16}{r_0^2}+\frac{74\alpha \eta }{3  r_0^4} \right)(r-r_-)(r-r_+)\,,\nn\\
f &=&\left(-\frac{12 r}{r_0^3} +\frac{16}{r_0^2}+\frac{2 \alpha  (37 \eta +64)}{3 r_0^4} \right) (r-r_+)(r-r_-)\,,\nn\\
W&=&6+\frac{2 \alpha  (13 \eta -8)}{3 r_0^2}-\frac{4 r}{r_0}\,,\nn\\
\omega &=&\frac{\sqrt{2} \alpha  (5 \eta +16)}{3 r_0^3}+\frac{3 \sqrt{2}}{r_0}-\frac{2 \sqrt{2} r}{r_0^2}\,.
\eea
The remaining thermodynamic quantities in this near extremal region are
\bea
T_\pm &=& \pm \frac{\sqrt{2\alpha  \eta }}{\pi  r_0^2}-\frac{2 \sqrt{2} \alpha  \eta }{\pi  r_0^3}\,,\qquad
\Omega_\pm = \frac{\sqrt{2}}{r_0}\mp\frac{2\sqrt{2\alpha  \eta }}{r_0^2}+\frac{4 \sqrt{2} \alpha  \eta }{3 r_0^3}\,,\nn\\
S_\pm &=& \pi^2 \Big(\frac{r_0^3}{\sqrt{2}}\pm \sqrt{2\alpha  \eta } r_0^2+\sqrt{2} \alpha  (6+\eta) r_0\Big).
\eea
It is easy to verify that the first law holds, up to and including the $\alpha$'th order. Thus we see that the near extremal blackening parameter $\eta$ must satisfy $\alpha\eta\ge 0$, with the extremal solution saturates the bound, in which case, the inner and outer thermodynamic variables coalesce.  Note that we have
\be
S_+ S_- = 4 \pi ^2 J^2 + 48 \pi ^{8/3} \alpha J^{4/3}\,.
\ee
Thus we see that at the $\alpha$ order, $S_+S_-$ is independent of the non-extremal parameter $\eta$ and hence is also quantized by the angular momentum, generalizing the results in two-derivative supergravities \cite{Cvetic:1996kv,Larsen:1997ge}.

\subsection{Gravitation, centrifugal and electrostatic forces}

For the extremal solution, we have
\be
M^{\rm ext} = \ft{3}{2} {\pi }^{\fft13} J^{\fft23} +\pi  \alpha\,,\qquad S^{\rm ext}=2\pi J + 12 \alpha \pi^{\fft53} J^{\fft13}\,.
\ee
We therefore obtain
\bea
\Delta_0 S &=& 12 \alpha \pi^{\fft53} J^{\fft13}+ {\cal O}(\alpha^2)\,,\nn\\
\Delta_\pm S &=&\mp 2 \sqrt{2} \pi ^{\fft43} \sqrt{\pm\alpha } J^{\fft23}+ 10 \pi ^{\fft53} \alpha  J^{\fft13} + {\cal O}(\alpha^2).
\eea
Under the unitarity of Gauss-Bonnet gravity $\alpha>0$, $\Delta_+ S$ is selected, and in addition
\be
\Delta_0S>0\,,\quad \Delta_{+}S<0\,.
\ee
Furthermore, the subadditivity (\ref{subadd}) is further strengthened. Compared to the RN result, it appears to be universal that the Gauss-Bonnet term with positive $\alpha$ strengthen the mass-charge relation towards subadditivity. This reaffirms the earlier observation that the unitary combinations of higher-order terms involving the curvature tensor tend to make gravity stronger.

The above inequalities suggest that the validity of the WGC requires to exclude the centrifugal force. Since electric charges can arise from the angular momenta in higher dimensions in the Kaluza-Klein perspective, this strong statement clearly deserves future investigation from point of view of effective field theory. Furthermore, the thermodynamical properties appear to be very similar between the RN and rotating black holes and we therefore would like to make a comparison. In appendix \ref{app:stability}, we review the thermodynamic instability analysis for both the RN and Kerr black holes in five dimensions.  For both black holes, one can consider grand canonical ensemble whose Gibb's free energy depends on temperature $T$ and $\Phi$ (or $\Omega$).  For both solutions, the specific heat $C_\Phi$ or $C_\Omega$ at fixed electric potential $\Phi$ or angular velocity $\Omega$ are negative for the whole black hole parameter space, indicating both black holes are thermodynamically unstable in the grand canonical ensemble.

For the canonical ensemble whose Helmholtz free energy depends on temperature $T$ and $Q$ (or $J$).  There is a critical mass $M^{\rm crit}(Q)$ or $M^{\rm crit}(J)$ such that when $M^{\rm ext} \le M<M^{\rm crit}$, the specific heat $C_Q$ or $C_J$ are both positive, and they change sign to become negative when $M> M^{\rm crit}$.  However this does not imply that the system is thermodynamically stable when $M<M^{\rm crit}$, since the charge capacitance $\widetilde C_T$ and ``angular momentum capacitance'' $\widehat C_T$ both have the opposite sign to the relevant specific heat, namely \cite{Monteiro:2008wr,Monteiro:2009tc}
\be
{\rm RN:}\qquad C_Q \widetilde C_T<0\,;\qquad\qquad {\rm Kerr:}\qquad C_J\widehat C_T<0\,.
\ee
Since the products are of the finite order, perturbative corrections can not reverse their sign. This implies that both RN and Kerr black holes are thermodynamically unstable in the canonical ensemble as well and their instabilities are parallel.

One can break this parallelity if one adopts the QSR and use the Euclidean action to compute the free energy.  The Euclidean actions of the RN and Kerr black hole both give the Gibb's free energy of grand canonical ensemble \cite{Gibbons:1976ue}.  However, for the Maxwell field, one can add a surface term
\be
I_{\rm Legendre}= \fft{1}{16\pi} \int_{\partial M} d\Sigma_\mu F^{\mu\nu} A_\nu\,,
\ee
which amounts to a Legendre transformation on the thermodynamical conjugate pair $(\Phi,Q)$ \cite{Gibbons:1976ue}.  However, there is no such a covariant term for the conjugate pair $(\Omega, J)$. This approach was indeed adopted in \cite{Camanho:2014apa,Cheung:2019cwi}, with the free energy given by
\be
F=M- T S - \Omega J\,,\qquad dF = -S dT - J d\Omega + \Phi dQ\,.
\ee
This in itself will not make the system stable since $C_Q \widetilde C_T$ is still negative.  In \cite{Camanho:2014apa,Cheung:2019cwi}, the stability is achieved by fixing the charge $Q$ literally such that $Q$ ceases to be a thermodynamic variable.  This leads to an approach that is in favour of the electric charge but discriminating against the angular momentum.  We here shall not judge on the validity whether this does solve the thermodynamic instability of the RN black hole in the canonical ensemble at the quantum level, but remark that the same trick cannot be applied for the Kerr black hole in the Euclidean action approach.

\section{Conclusion}

In this paper we studied the RN black holes perturbed by higher-derivative invariants. We studied the relations among the mass, charge and entropy in the extremal limit. For fixed charge, extremal black holes continue to exist and
the change of the mass $\Delta M^{\rm ext}$ due to the higher order correction is proportional to the coupling constants. The mass and entropy differences with the unperturbed extremal solution satisfy
\be
\Delta M^{\rm ext}(Q) \propto + \Delta S^{\rm ext}(Q)\,.
\ee
In order to obtain the entropy difference for fixed mass and charge, we have to move away from extremality since the perturbed and original black hole cannot be extremal simultaneously. We focus on the near extremal region with the blackening factor of the order of the coupling constant, namely
\be
M=M^{\rm ext} + (1 + \eta) \Delta M^{\rm ext}.
\ee
where $\eta$ is the dimensionless non-extremal parameter of order 1.   We found in this near extremal region that
\be
\Delta M^{\rm ext}(Q) \propto - \sqrt{|\alpha|} \Delta S(\eta,Q)\,,
\ee
Here $\alpha$ represents the coupling constant or some certain linear combination of the coupling constants.  Imposing causality, unitarity and analyticity, the higher-order corrections of the pure matter sector are not only indispensable and but also must be dominant over the sectors involving gravity for the WGC. Based on our observation, it seems to be sensible to conjecture that well-defined higher-order corrections involving curvature invariants tend to make gravity stronger, whilst the higher-order pure matter sector satisfying the appropriate energy conditions tends to make gravity weaker. If this conjecture can be shown generally true, the WGC would imply that a UV completed quantum gravity necessarily involve matter.

Our main focus was to construct the five dimensional rotating black hole in pure Einstein gravity extended with quadratic curvature invariants, up to and including the linear order of the coupling constant.  We expect that it is consistent to truncate out all the matter field so that we can focus on a smaller but self consistent effective of pure quantum gravity. Since the background solution is Ricci flat, it follows that $R^2$ an $R^{\mu\nu} R_{\mu\nu}$ terms do not contribute at the linear order. Therefore the solution is also for the Einstein-Gauss-Bonnet theory, with the standard coupling constant $\alpha$.  In five dimensions, the rotating black hole metric becomes cohomogeneity one, depending on only the radial coordinate, when the two angular momenta become equal.  This allows us to obtain the linearly perturbed analytic solution for general mass and (equal) angular momenta. We analysed the geometric and thermodynamic properties of the black hole. The general solution has only one horizon. The inner horizon emerges when the solution is sufficiently close to extremality, at the order of the coupling constant.  We found that the entropy product of the inner and outer horizons in this case depends only on the angular momentum, which is expected to quantized at the quantum level.

Our main conclusion is that for solutions close to extremality, for positive $\alpha$, we have $\Delta S<0$ and furthermore in the extremal limit, $M^{\rm ext} = \ft{3}{2} {\pi }^{\fft13} J^{\fft23} +\pi  \alpha$.  The positive sign of the $\alpha$-correction implies that the centrifugal repulsion associated with rotations becomes even weaker than the gravitational attraction under the unitary requirement for the Gauss-Bonnet term.  The subadditivity of $M(J)$ is further strengthened by the Gauss-Bonnet correction. Our results suggest that the WGC has to exclude the consideration of the centrifugal repulsion, which is stronger in classical gravity and becomes even stronger in the quantum corrected theories. It is thus of great interest to investigate how this statement is consistent with the Kaluza-Klein perspective, where the angular momenta are reduced to become electric charges in lower dimensions.

Finally we observe that in the case of Einstein-Maxwell theory, the WGC suggests that after the quantum correction, the linear mass/charge relation is modified such that the extremal black holes are energetically favoured to split up. On the other hand, it is well established in the semi-classical limit that the RN black holes are entropically favoured to join together \cite{Cvetic:2018dqf}. These seemingly contradictory tendencies require a resolution.

\section*{Acknowledgement}

We are grateful to Jun-Bao Wu and Yi Pang for useful and extensive discussions. The work is supported in part by NSFC (National Natural Science Foundation of China) Grants No.~11875200 and No.~11935009.

\section*{Appendix}
\appendix

\section{The general perturbative solution}
\label{app:sol}

The linear perturbation (\ref{rotpert}) of the five dimensional rotating black hole with equal angular momenta can be solved exactly and the general solutions involve six integration constant $c_i$, $i=1,\ldots,6$, (not to be confused with the coupling constants in section 2.)  The metric functions are
\bea
\Delta f &=& \frac{240 \mu  \nu ^2 -3 c_3 \tilde{\nu }^5+\frac{32 \mu ^5}{\nu ^2}-188 \mu ^3}{\tilde{\nu }^4}-r^2 \left(\frac{48 \mu ^2}{\tilde{\nu }^4}+c_2\right)\nn\\
&&+\frac{c_1-\frac{4 \left(16 \mu ^6-79 \mu ^4 \nu ^2+72 \mu ^2 \nu ^4\right)}{\nu ^2 \tilde{\nu }^4}}{r^2}+2 c_2 \mu -\frac{16 \nu ^4}{r^{10}}+\frac{64 \mu  \nu ^2}{3 r^8}+\frac{40 \mu ^2}{3 r^6}\nn\\
&&+\frac{64 \mu ^3}{3 \nu ^2 r^4}-\frac{\left(2 \mu ^2+\nu ^2+3 r^4-6 \mu  r^2\right)}{6 \nu ^4 r^2}(P_++P_-)\,,\cr
\Delta h &=&c_4 \Big(1 - \fft{2\mu}{r^2} + \fft{\nu^2}{r^4}\Big) +
\frac{3 c_3 \nu ^2 \tilde{\nu }+\frac{12 \mu  \nu ^4}{\tilde{\nu }^4}+\frac{220 \mu  \nu ^2}{\tilde{\nu }^2}-\frac{64\mu \left(\mu ^2+2  \nu ^2\right)}{3 \nu ^2}}{r^4}\nn\\
&&-\frac{\frac{c_3 \tilde{\nu } \left(5 \mu ^2+\nu ^2\right)}{\mu }+\frac{6 \nu ^2 \left(\mu ^2+\nu ^2\right)}{\tilde{\nu }^4}+\frac{254 \nu ^2}{\tilde{\nu }^2}-c_2 \nu ^2-c_1-\frac{64 \mu ^2}{3 \nu ^2}+168}{r^2}\cr
&&-\frac{16 \nu ^4}{3 r^{10}}+\frac{32 \mu  \nu ^2}{3 r^8}+\frac{8 \left(9 \mu ^2+4 \nu ^2\right)}{3 r^6}\nn\\
&&-\frac{(2\mu^2 + \nu^2)(\nu^2 +r^4)-2 \mu r^2\left(\mu ^2+2  \nu ^2\right)}{6 \mu  \nu ^4 r^4} (P_+ + P_-)\,,\nn\\
\Delta W &=& \fft{c_5}{r^4} +\frac{3 c_3 \nu^2 \tilde{\nu }^5-c_2 \mu  \nu ^2 \tilde\nu^4-4 \left(8 \mu ^5-59 \mu ^3 \nu ^2+60 \mu  \nu ^4\right)}{2 \nu ^2 \tilde{\nu }^4}\nn\\
&&+\frac{r^2 \left(c_2 \tilde \nu^4 +48 \mu ^2\right)}{2 \tilde{\nu }^4}+\frac{16 \mu ^2 \left(2 \mu ^4-4 \mu ^2 \nu ^2-\nu ^4\right)-c_2 \nu ^4 \tilde\nu^4}{2 \nu ^2\tilde{\nu }^4 r^2}\nn\\
&&+\frac{16 \nu ^4}{3 r^{10}}-\frac{16 \mu  \nu ^2}{3 r^8}-\frac{16 \left(\mu ^2+3 \nu ^2\right)}{3 r^6}\nn\\
&&
+\frac{\mu ^2 \nu ^2+2 \nu ^4+3 \mu  r^6-3 \mu ^2 r^4-3 \mu  \nu ^2 r^2}{12 \mu  \nu ^4 r^4}(P_++P_-)\,,\nn\\
\Delta \omega &=&c_6 + \frac{1}{3 \sqrt{2} \mu ^{3/2} \nu  \tilde{\nu }^4 \left(\nu ^2+r^4\right)r^6}
\Bigg(3 c_3 \nu^2\tilde{\nu }^5 \left(2 \mu ^2+\nu ^2+3 \mu  r^2\right)r^6\nn\\
&&+\tilde \nu^4 \Big(6c_5 \mu^2 r^6 - 64\mu^2 (\mu^2 r^4 + \mu\nu^2 r^2 - \nu^4)\Big)\nn\\
&&-3 \mu  r^6 \Big(\nu^2 \tilde \nu^4 \left(c_1 +c_2 (2 \mu  r^2-\tilde \nu^2)-2c_4\mu\right)\cr
&&+4 \left(-6 \mu ^6+9 \mu ^4 \nu ^2-31 \mu ^2 \nu ^4+16 \nu ^6+\mu r^2\left(8 \mu ^4-47 \mu ^2 \nu ^2+60 \nu ^4\right)\right)\Big)\Bigg)\nn\\
&&+\frac{2 \mu ^2+\nu ^2+3 r^4-6 \mu  r^2}{6 \sqrt{2} \sqrt{\mu } \nu ^3\left(\nu ^2+r^4\right)} (P_+ + P_-)\,,\nn
\eea
where
\bea
P_\pm  &=& \Big(\pm \fft{4 \mu }{\tilde\nu^5} \left(16 \mu ^6-40 \mu ^4 \nu ^2+87 \mu ^2 \nu ^4-60 \nu ^6\right)\pm 3 c_3 \nu ^4 - 64 \mu ^2\Big)\log \left(1-\frac{\mu \pm \tilde \nu }{r^2}\right)\,.
\eea
In order for the solution not to generate a cosmological constant, we set
\be
c_2=-\frac{48 \mu ^2}{\tilde \nu^4}\,.
\ee
For the solution to be asymptotically Minkowski in the static frame with $g_{tt}=-1$, we set $c_4=0=c_6$.
We further impose that the mass and angular momentum remain fixed under the linear $\alpha$ perturbation, which implies
\bea
c_1 &=&\frac{16 \mu ^2 \left(2 \mu ^2+\nu ^2\right)}{\tilde{\nu }^4}+3 c_3 \mu  \tilde{\nu }\,,\nn\\
c_5 &=&-\frac{c_3 \nu ^2 \tilde{\nu } \left(2 \mu ^2+\nu ^2\right)}{2 \mu ^2}+\frac{2 \left(10 \mu ^6-58 \mu ^4 \nu ^2+29 \mu ^2 \nu ^4+16 \nu ^6\right)}{\mu  \tilde{\nu }^4}\,.
\eea
The general solution is singular at $r=r_\pm^0$, given by (\ref{bghorizons}).  We can choose the parameter $c_3$ such that the solution is regular at either $r_+^0$ or $r_-^0$:
\bea
c_3=C_+\,,&&\qquad \hbox{preserving regularity at $r=r_+^0$,}\nn\\
c_3=C_-\,,&&\qquad \hbox{preserving regularity at $r=r_-^0$,}
\eea
where
\be
C_\pm=\pm  \frac{64 \mu ^2}{3 \nu ^4}-\frac{4 \mu  \left(16 \mu ^6-40 \mu ^4 \nu ^2+87 \mu ^2 \nu ^4-60 \nu ^6\right)}{3 \nu ^4 \tilde{\nu }^5}\,.
\ee
In section \ref{sec:rot}, we gave the $c_3=C_+$ solution, in which case, the $r=r_0^-$ is the spacetime singularity, outer of which there can either one, two or three horizons depending on the sign of the coupling constant $\alpha$ and its relative size to the non-extremal factor $\tilde \nu$.
Here we present the $c_3=C_-$ solution:
\bea
\Delta h &=&-\frac{8}{3 \nu ^4 r^{10}}\Big(\mu  \nu ^4 r^4 \left(4 r^2-9 \mu \right) -4 \nu ^6 r^2 \left(\mu +r^2\right)+2 \nu ^8-16 \mu ^3 r^8 \left(\tilde{\nu }+\mu \right)\nn\\
&&-8 \mu  \nu ^2 r^6 \left(\tilde{\nu } \left(r^2-3 \mu \right)+\mu  \left(r^2-4 \mu \right)\right)
\Big)\nn\\
&&+\frac{64 \mu  \left((2\mu^2+ \nu^2) (\nu^2 + r^4) - 2 \mu r^2 (\mu^2 + 2\nu^2)\right)}{3 \nu ^4 r^4}\log \left(1-\frac{\tilde{\nu }+\mu }{r^2}\right),\cr
\Delta f &=& \frac{8 \left(24 \mu ^2 r^8 \left(r^2-\mu \right) \left(\tilde{\nu }+\mu \right)-6 \nu ^8+5 \mu ^2 \nu ^4 r^4+8 \mu  \nu ^6 r^2+4 \mu ^2 \nu ^2 r^6 \left(2 \mu -3 r^2\right)\right)}{3 \nu ^4 r^{10}}\nn\\
&&+\frac{64 \mu ^2 \left(2 \mu ^2+\nu ^2+3 r^4-6 \mu  r^2\right) }{3 \nu ^4 r^2}\log \left(1-\frac{\tilde{\nu }+\mu }{r^2}\right),\nn\\
\Delta W &=& -\frac{4}{3 \mu  \nu ^4 r^{10}}\Big(6 \nu ^6 r^6+8 \mu ^4 r^6 \left(3 r^4-2 \nu ^2\right)+\mu ^2\nu^4 r^2 \left(4 \nu ^2-15 r^4\right)\nn\\
&&+4 \mu ^3 r^4 \left(6 \tilde{\nu }r^6-\nu ^2 r^2\left(4\tilde{\nu }+3 r^2\right)+\nu ^4\right)-4 \mu \nu^4 \left(2\tilde{\nu }r^6+\nu ^4-3 \nu ^2 r^4\right)\Big)\nn\\
&&-\frac{32 \mu  \left(2 \nu ^4+\mu ^2 \left(\nu ^2-3 r^4\right)+3 \mu  r^2\left(r^4-\nu ^2\right)\right) }{3 \nu ^4 r^4}\log \left(1-\frac{\tilde{\nu }+\mu }{r^2}\right),\nn\\
\Delta \omega &=& -\frac{16 \sqrt{2} \sqrt{\mu } \left(6 \mu  r^6 \left(r^2-\mu \right) \left(\tilde{\nu }+\mu \right)-2 \nu ^6+2 \mu  \nu ^4 r^2+\mu  \nu ^2 r^4 \left(2 \mu -3 r^2\right)\right)}{3 \nu ^3 r^6 \left(\nu ^2+r^4\right)}\nn\\
&&-\frac{32 \sqrt{2} \mu ^{3/2} \left(2 \mu ^2+\nu ^2+3 r^4-6 \mu  r^2\right) }{3 \nu ^3 \left(\nu ^2+r^4\right)}
\log \left(1-\frac{\tilde{\nu }+\mu }{r^2}\right).
\eea
The curvature singularity of the perturbed solution is now located at $r=r_+^0$.  It can easily established that when $\alpha>0$, there are no roots for both functions $h$ and $f$ at $r>r_+^0$.  The singularity is thus naked.  When $\alpha<0$, both $h$ and $f$ can have roots at $r>r_+^0$, but they do not coincide. However, the root of $f$ is larger than $h$, indicating the solution is a rotating wormhole. We shall not study this solution in detail, but present an example: $(\mu,\nu,\alpha)=(5,3,-1/1000)$, we have $r_+^0=3$ and the roots of $h$ and $f$ are given by
\be
h(3.00540)\sim 0\,,\qquad f(3.00775)\sim 0\,.
\ee

\section{Thermodynamic instability of RN and Kerr black holes}
\label{app:stability}

In this section, we review the thermodynamic stability of the RN and Kerr black holes in five dimensions, by examining the specific heat $C$, charge capacitance $\widetilde C$ and ``angular momentum capacitance'' $\widehat C$ in both canonical and grand canonical ensembles.  Although these results are well studied in literature, {\it e.g.}, \cite{Monteiro:2008wr,Monteiro:2009tc}, it is nevertheless instructive to put them together to see the parallelity.

   First we consider the RN black hole of mass $M$ and charge $Q$, given in section \ref{sec:EMGB}. We have
\bea
&&T=\frac{\sqrt{\frac{3}{\pi }} \sqrt{M^2-3 Q^2}}{2 \left(\sqrt{M^2-3 Q^2}+M\right)^{3/2}}\,,\qquad
S=\frac{4}{3} \sqrt{\frac{\pi }{3}} \left(\sqrt{M^2-3 Q^2}+M\right)^{3/2}\,,\nn\\
&& \Phi=\frac{3 Q}{\sqrt{M^2-3 Q^2}+M}\,.
\eea
In the grand canonical ensemble of fixed $T$ and $\Phi$, the thermodynamic potential is Gibb's free energy $G(T,\Phi)=M-T S - \Phi Q$ and we can define the specific heat $C$ and charge capacitance $\widetilde C$
\bea
C_\Phi &=& T \fft{\partial  S}{\partial T}\Big|_\Phi =-4 \sqrt{\frac{\pi }{3}} \left(\sqrt{M^2-3 Q^2}+M\right)^{3/2}\,,\nn\\
\widetilde C_T &=& \fft{\partial Q}{\partial \Phi}\Big|_T=\frac{M \left(\sqrt{M^2-3 Q^2}+M\right)-9 Q^2}{3 \sqrt{M^2-3 Q^2}}\,.\label{CPHICT}
\eea
Thus we see that the specific heat is always negative in this ensemble, indicating the solution is not thermodynamically stable.  There is a phase transition occurring at
\be
M^{\rm crit}=3 \sqrt{\frac{3}{5}} Q\,,
\ee
at which $\widetilde  C_T$ vanishes.  For the canonical ensemble of fixed $T$ and $Q$, the corresponding thermodynamic potential is the Helmholtz free energy $F(T,Q)=M-T S$.  In this case, the specific heat at fixed charge is
\be
C_Q = T \fft{\partial  S}{\partial T}\Big|_Q= -\frac{8 \pi  T\left(\sqrt{M^2-3 Q^2}+M\right)^4}{3 \left(M \sqrt{M^2-3 Q^2}+M^2-9 Q^2\right)}\,.
\ee
The charge capacitance at fixed temperature $\widetilde C_T$ is the same as in (\ref{CPHICT}).  Thus we see that \cite{Monteiro:2008wr}
\be
C_Q \widetilde C_T = -\ft{4}{3} \sqrt{\ft{\pi }{3}} \sqrt{\sqrt{M^2-3 Q^2}+M}\Big(2 M \sqrt{M^2-3 Q^2}+2 M^2-3 Q^2\Big)<0.
\ee
Note that there is a phase transition at $M^{\rm crit}$ where both $C_Q$ and $\widetilde C_T$ change the sign. Thus we see that the RN black hole is thermodynamically unstable in both ensembles.

We now consider the five dimensional rotating metric, with equal angular momenta.  For given $(M,J)$, we have
\bea
&&T=\frac{2 M^2 \sqrt{16 M^2-\frac{54 \pi  J^2}{M}}+27 \pi  J^2-8 M^3}{18 \sqrt{6} \pi ^{3/2} J^2 \sqrt{M}}\,,\nn\\
&&
S=\ft{1}{3} \sqrt{\ft{2}{3}\pi M} \left(\sqrt{16 M^2-\frac{54 \pi  J^2}{M}}+4 M\right)\,,\qquad \Omega=\frac{4 M-\sqrt{16 M^2-\frac{54 \pi  J^2}{M}}}{6 J}\,.
\eea
Thus for the grand canonical ensemble with fixed $(T,\Omega)$, associated with the Gibb's free energy, the specific heat is
\bea
C_\Omega &=& T \fft{\partial  S}{\partial T}\Big|_\Omega\nn\\
&=&- \fft{T}{9M^{\fft52}}\Big(16 \left(6 M^3-M_{\rm ext}^3\right) \sqrt{M^3-M_{\rm ext}^3}+32 M^{3/2} \left(3 M^3-2 M_{\rm ext}^3\right)\Big).
\eea
This quantity is always negative since we must have $M\ge M^{\rm ext}$, where $M^{\rm ext}$ is given by (\ref{kerrmassext}). This is exactly parallel to the RN black hole. In the canonical ensemble with fixed $(T,J)$, associated with the Helmholtz free energy, the specific heat is
\bea
C_J &=& T \fft{\partial  S}{\partial T}\Big|_J\nn\\
&=& \fft{4\sqrt{\ft{2 \pi }{3}} M^{\fft32} M_{\rm ext}^3 \sqrt{M^3-M_{\rm ext}^3}}{M^{\fft32}(5M^3-2M_{\rm ext}^3)-
(5M^3+M_{\rm ext}^3) \sqrt{M^3-M_{\rm ext}^3}}\,.
\eea
There is a critical mass
\be
M^{\rm crit} = \Big(\ft{1}{10} \left(3 \sqrt{21}+13\right)\Big)^{\fft13} M^{\rm ext}\,,
\ee
below which $C_J$ is positive, but it changes to negative when $M\ge M^{\rm crit}$. The angular momentum capacitance $\widehat C$ at fixed $T$ (which is called the isothermal moment of inertia in \cite{Monteiro:2009tc}) is the same for both ensembles, and we have
\bea
\widehat C_{T} &=& \fft{\partial J}{\partial \Omega}\Big|_T\nn\\
&=& \fft{4M_{\rm ext}^3\left(\left(M^3-M_{\rm ext}^3\right) \left(5 M^3+3M_{\rm ext}^3\right)+M^{3/2} \left(2 M_{\rm ext}^3-5 M^3\right) \sqrt{M^3-M_{\rm ext}^3}\right)}{9\pi  M^{\fft52}\left(\left(2 M^3-M_{\rm ext}^3\right) \sqrt{M^3-M_{\rm ext}^3}+2 M^{3/2} \left(M_{\rm ext}^3-M^3\right)
\right)}.
\eea
This leads to \cite{Monteiro:2009tc}
\be
C_Q \widehat C_{T}=-\fft{64 T}{27\sqrt{M}}\Big(\left(4 M^3-M_{\rm ext}^3\right) \sqrt{M^3-M_{\rm ext}^3}+M^{3/2} \left(4 M^3-3 M_{\rm ext}^3\right)\Big)<0\,.
\ee
As in the RN case, the rotating solution is also thermodynamically unstable in both ensembles.  However, in the Euclidean action approach, one can break this parallelity, which we discuss at the end of section \ref{sec:rot}.

\end{document}